\documentclass[a4paper,11pt]{article}
\usepackage{pos}
\usepackage{graphicx}
\usepackage{wrapfig}
\begin{document}

\title{Abundant radiation of soft photons:\\
a puzzle lasting four decades}
\ShortTitle{Soft photons}

\author*{Boris Kopeliovich}
\author{Irina Potashnikova}
\author{ Iv\'an Schmidt}

\affiliation{Departamento de F\'{\i}sica,
Universidad T\'ecnica Federico Santa Mar\'{\i}a,\\
 Avenida Espa\~na 1680, Valpara\'iso, Chile}




\emailAdd{boris.kopeliovich@usm.cl}
\emailAdd{irina.potashnikova@usm.cl}
\emailAdd{ivan.schmidt@usm.cl}

\abstract{The observed enhancement of low-$k_T$ photons in comparison with incorrect calculations, should not be treated as a puzzle.
The paper by Low considered a large rapidity gap process of diffractive excitation of a hadron, $h\to h+\gamma$, rather than  multiple hadron production spanning all over the rapidity interval between colliding hadrons. The optical theorem connects these two processes, and what is inner bremsstrahlung, suppressed according to Low, corresponds to radiation from final state hadrons. Thus, the main result of the Low theorem, based on gauge invariance of the diffractive bremsstrahlung amplitude, supplemented with the optical theorem, contradicts the so-called bremsstrahlung model. The latter has been used for comparison with data, leading to the longstanding soft photon puzzle.}

\FullConference{16th International Symposium on Radiative Corrections: Applications of Quantum Field Theory to Phenomenology (
RADCOR2023)\\
28th May - 2nd June, 2023\\
Crieff, Scotland, UK\\}


\maketitle

\section{The Low theorem revisited}
\begin{figure}[h]
    \centering
    \includegraphics[scale=0.25]{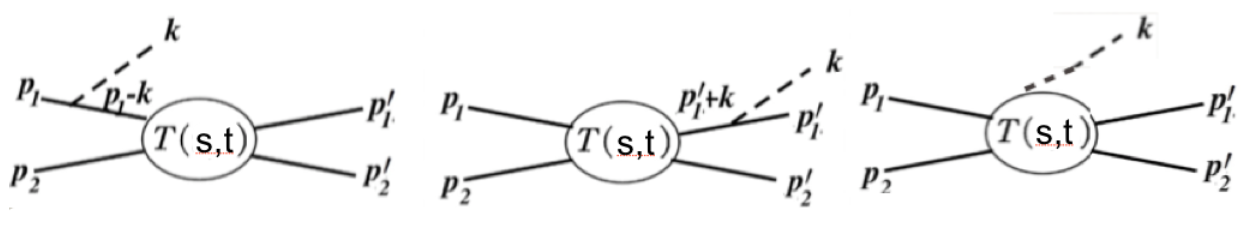}
    \caption{External (1st and 2d figs.) and internal (3rd fig.) terms in the radiation amplitude.}
    \label{fig:3graphs}
\end{figure}

The process under consideration in the Low paper \cite{Low} is quasi-elastic, hadronic scattering with radiative excitation of one of the hadrons,
\begin{equation}
    h_1+h_2\to h_1^\prime+\gamma +h_2^\prime
    \label{h1-h2}
\end{equation}
The mechanisms of radiation can be grouped to external and internal radiation as is depicted
in Fig.~\ref{fig:3graphs}.

Correspondingly, the amplitude of (\ref{h1-h2}) reads, $ M=e_{\mu} M_\mu,$,
where  $e_\mu$ is the photon polarization vector, and
\begin{equation}
M_\mu=M_\mu^{ext}+M_\mu^{int}
\label{M-mu}
\end{equation}

The amplitudes of external radiation contains Feynman propagators,
\begin{equation}
    M_\mu^{ext}=\left(\frac{p_{1\mu}^\prime}{p_1^\prime k}
    -\frac{p_{1\mu}}{p_1 k}\right) T(s,t)+...
\label{M-ext}
\end{equation}
The factor in parentheses is divergent at $k\to0$.
The elastic hadronic amplitude $h_1h_2\to h_1^\prime h_2^\prime$ is depicted in Fig.~\ref{fig:3graphs} by blobs.

The two terms in Eq.~(\ref{M-mu}) are related since the amplitude is Gauge invariant \cite{Low}, $k_\mu M_\mu = 0$. Therefore, the term presenting internal radiation, must be finite at $k\to0$, i.e. is suppressed in comparison with external radiation. This is the main observation of the Low theorem \cite{Low}. 

\subsection{Landau-Pomeranchuk principle}

The Low theorem can be also treated as a formal proof of the Landau-Pomeranchuk principle \cite{lp}, which states that any variations of the electric current within a short distance do not affect the spectrum of radiation with a much longer radiation, or coherence length,
\begin{equation}
    l_c^\gamma=\frac{2E_{h}x(1-x)}{k_T^2+x m_h^2},
    \label{lc}
\end{equation}
where $x=k_+^\gamma/p_+^h$ is the fractional light-cone hadronic momentum carried away by the photon. The important condition for the Low theorem is a long radiation coherence length, which Low calls “the distance a particle can move with energy imbalance $\Delta E=k$”. Notice that $l_c^\gamma$ and $E_{h1}$  are not Lorentz invariant, so must be taken within the same reference frame.

Important is to keep the incoming ($l<-l^\gamma_c$) and 
outgoing ($l>l^\gamma_c$)  currents unaffected by the current variations on a short length scale of strong interaction. This means 
that only extrinsic radiation from initial and final hadrons $h_1$, $h_1^\prime$ matters.

\subsection{Fock state representation}

According to the optical theorem forward elastic amplitude is a "shadow" of inelastic interactions. This is why the elastic process is usually called diagonal diffraction.  Good and Walker \cite{gw} proposed an extended version of the optical theorem, 
relating the off-diagonal diffractive amplitude with a linear combination of diagonal amplitudes. This can be explained basing on the Fock-state expansion of the hadronic wave function. For radiactive diffractive excitation for the sake of simplicity we single out two Fock components of the hadron, just a hadron $|h\rangle_0$, and a hadron accompanied by a photon $|h\gamma\rangle$ \cite{kst2}, 
\begin{equation}
    |h\rangle=C_0|h\rangle_0+C_1|h\gamma\rangle+ ...
\label{fock}
\end{equation}
The Fock components are eigenstates of the elastic amplitude operator, $\hat f|i\rangle = f_i|i\rangle$.

Diffractive excitation occurs only due to diversity of the elastic eigenamplitudes, otherwise the incoming wave packet would remain unchanged, so only elastic scattering was possible. Indeed, according to (\ref{fock}) the off-diagonal diffractive  amplitudes read,
\begin{equation}
    \langle h\gamma|\hat f|h\rangle = C_1^* C_0 (f_{h\gamma}-f_h)
\label{offdiagonall}
\end{equation}
Differently from the Feynman diagrams, having no space-time structure, one cannot say whether the photon is radiated before or after the interaction. The radiation amplitude is a linear combination of elastic eigenamplitudes.

Notice that in both Fock components $|h\rangle$ and $|h \gamma\rangle$ only the hadron constituent $h$ interacts, so at first glance 
$f_{h} = f_{h\gamma}$ and the diffractive amplitude must vanish. Nevertheless, the impact parameters b are shifted by the radiation, so
the eigen-amplitudes $f_{h}(b) \neq f_{h\gamma}(b^\prime)$ cancel only in the $b$-integrated amplitude, i.e. at $t=0$. Indeed, the Low amplitude Eq.~(\ref{M-ext}) vanishes in forward direction, $t=0$, what corresponds to integration of the amplitude over all impact parameters.

\section{Photon radiation in inelastic collisions}

\subsection{Use and misuse of the Low theorem}

The so-called bremsstrahlung model (BM) pretends to extend the Low theorem to inelastic collisions with multi-particle production. 
 Photons are assumed to be radiated by participating
charge particles, either incoming, or outgoing.

\begin{wrapfigure}{l}{0.3\textwidth}
\centering
    \includegraphics[width=4cm]{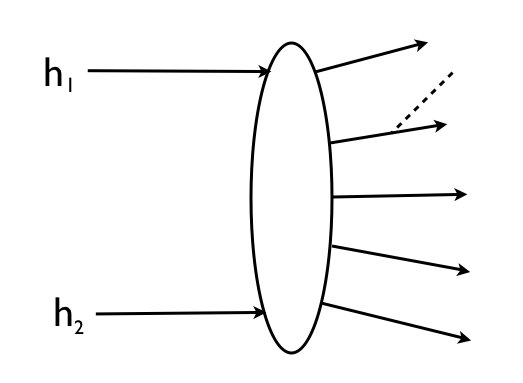}
    \caption{Multi-hadron production and photon radiation in inelastic collision.}
    \label{fig:inelastic}
\end{wrapfigure}
    It might be tempting to generalize the Low expression (\ref{M-ext})  
    to a larger number of produced charged hadrons, as was  proposed in \cite{goshaw} (see more references in \cite{wong}). 
\begin{equation}
M=M_0(p_1,p_2;p_3...p_N)\left(\sum\limits_i 
\frac{\eta_ie_ip_i\cdot\epsilon}{2p_i\cdot k}\right),
\label{in}
\end{equation}
where $\eta_i=\pm 1$ for outgoing and incoming particles respectively, and $M_0(p_1,p_2;p_3...p_N)$ is the amplitude of $2\to N$ inelastic collision without radiation.
Such an unjustified extension of the Low result derived for a diffractive process,  is illegitimate  \cite{icnfp}. Moreover, it strictly contradicts the Low theorem, as is demonstrated below. Not a surprise that BM contradicts data.

\subsection{Bremsstrahlung Model vs data}

\begin{figure}[h]
    \centering
    \includegraphics[width=15cm]{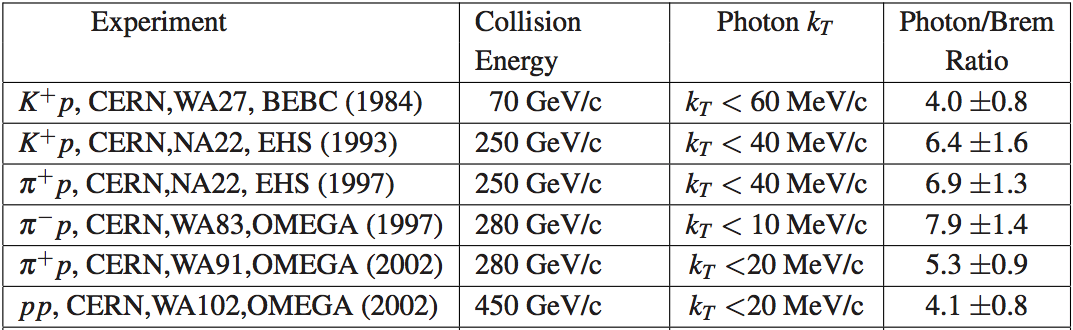}
    \caption{The list of experiments detected an anomalous enhancement of low-$k_T$ photon in high-energy inelastic hadronic collisions. The last column shows the ratio of the observed yield of soft photons to the prediction of the BM. The table is taken from \cite{wong}}.
    \label{fig:puzzle}
\end{figure}

A puzzling enhancement of low-$k_T$ photons in high-energy hadronic and hadron-nucleus collisions in comparison with BM expectations has been repeatedly observed since 1984. A collection of relevant experimental results \cite{wong}, is displayed in Fig.~\ref{fig:puzzle}.
Data are compared with predictions of the BM  Eq.~(\ref{in}). We see that the yield of low-$k_T$ photons substantially exceeds data. Apparently, the suspect is the model.

\subsection{The Low process at high energies}
 At high energies the Low process essentially simplifies. Diffraction is a large rapidity gap process. That means that the hadronic amplitude $T(s,t)$ is dominated by Pomeron exchange in terms of Regge phenomenology \cite{kaidalov}, or by colorless gluonic exchange \cite{bfkl} in terms of QCD.
 
 \begin{wrapfigure}{l}{0.45\textwidth}
     \centering
     \includegraphics[width=6cm]{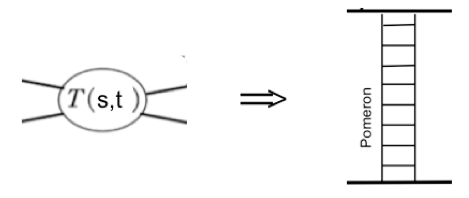}
     \caption{The hadronic amplitude $T(s,t)$ is dominated by the Pomeron exchange.}
     \label{fig:pomeron}
 \end{wrapfigure}
  The ladder graph Pomeron  depicted in Fig.~\ref{fig:pomeron} is simplified. In topological expansion the Pomeron has non-planar, or cylindrical structure \cite{kaidalov}.

According to Regge phenomenology the Regge intercept of the Pomeron trajectory is slightly above one, $\alpha_{Pom}=1+\Delta$, so the elastic amplitude $T(s,t)\propto s^\Delta$. Since $\Delta\ll 1$ the energy derivatives $\partial T(s,t)/\partial s\sim 1/s$ vanish. This makes the relation ~(\ref{M-ext}) more accurate, because the skipped derivative terms \cite{Low} can be neglected.


\subsection{Unitarity relation}

\begin{wrapfigure}{l}{0.4\textwidth}
\includegraphics[width=5cm]{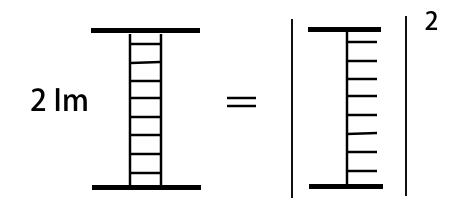}
\caption{Unitarity cut of the Pomeron.}
\label{fig:unitarity}
\end{wrapfigure}

Optical theorem relates the inelastic cross section for multi-particle production, with the imaginary part of the forward elastic amplitude, as is depicted in Fig.~\ref{fig:unitarity} for the Pomeron amplitude.
\newpage

\vspace{1.2cm}

Such a representation for the Pomeron 
 is usually called a cut Pomeron. 
For the sake of simplicity we neglect the 
small elastic cross section in the r.h.s. the relation 
Fig.~\ref{fig:unitarity}.

Unitarity relation for multi-Pomeron amplitudes (e.g. multiple interactions in nuclei) is given by the AGK cutting rules \cite{agk}, calculating the contribution  of a certain number of inelastic collision.

\subsection{Charge screening}

The specific feature of the Pomeron is electric neutrality. This means that electric charges of the colliding hadrons $h_{1,2}$ do not flow toward the central rapidities. A charge can be transported through large rapidity intervals only by Reggeons, having intercepts $\alpha_R\approx 1/2$ and lower. Therefore production of a charged hadron with rapidity $y$ is suppressed by $\exp(-(Y-y)/2$, where $Y=\ln(\sqrt{s}/2)$. Thus, the electric charges of the colliding hadrons remain at maximal and minimal rapidities and photons cannot be radiated at medium rapidities as is incorrectly assumed in the BM Eq.~(\ref{in}). 

Of course one could introduce quark loops in the gluonic latter of the Pomeron, however photon radiation from quarks and antiquarks cancels due to charge screening.

\subsection{Why Feynman rules cannot be applied to Fig.~\ref{fig:inelastic} }

The infra-red divergent Feynman propagator
$\eta_ie_ip_i\cdot\epsilon/2p_i\cdot k$
in (\ref{in}) in coordinate representation reads,
\begin{equation}
\frac{1}{p^2-m_h^2}=\frac{l_c^\gamma}{2E}
\end{equation}
When the propagator diverges as $1/k$ the coherence length rises to infinity. This means that a hadron radiating soft photons, cannot emerge instantaneously, but its production takes a long time.
\begin{wrapfigure}{l}{0.3\textwidth}
    \centering
    \includegraphics[width=4.5cm]{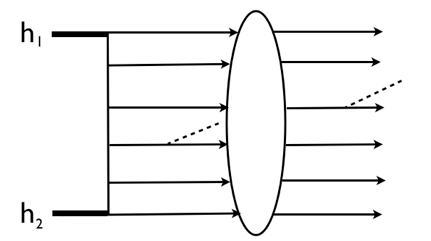}
    \caption{Space-time pattern of radiative inelastic interaction.}
    \label{fig:space-time}
\end{wrapfigure}
This is illustrated in Fig.~\ref{fig:space-time} which shows that all final hadrons radiating a photon must preexist long time before interaction. This is a standart space-time picture of interaction in the conventional parton model.

Thus, the process cannot be treated like in (\ref{in}) as radiation
either from the two incoming, or from N outgoing charges. All the produced partons pre-exist prior the interaction. Due to factorization of long distances (Partonic wave function of a hadron) and short-scale interaction the cross section is subject to the factorization theorem.
This inconsistency is another reason for incorrectness of the Bremsstrahlung Model.

\section{Color-dipole description of photon radiation}

 As far as the BM turns out to be incorrect, an alternative description of soft photon radiation is required \cite{modelling}.

The cross section of bremsstrahlung by a projectile quark can be calculated within the color dipole phenomenology 
\cite{zkl,BK1,kst1,kst2,krt,applied}, adjusted to precise data on DIS from HERA,
\begin{equation}
\frac{d\sigma(qN\to\gamma X)}{d\ln{\alpha} d^2k_T} =
\frac{1}{(2\pi)^2}\int d^2r_1 d^2r_2
e^{i\vec k_T(\vec r_1-\vec r_2)}
\Psi^*_{\gamma q}(\alpha,\vec r_1)\Psi_{\gamma q}(\alpha,\vec r_2)\,
\sigma_\gamma(\vec r_1,\vec r_2,\alpha),
\label{DM}
\end{equation}
where $\alpha$ is the fractional light-cone momentum, carried by the photon;
\begin{equation}
\sigma_\gamma(\vec r_1,\vec r_2,\alpha)=
\frac{1}{2}\left\{\sigma_{\bar qq}(\alpha r_1)+
\sigma_{\bar qq}(\alpha r_2)-
\sigma_{\bar qq}[\alpha(\vec r_1-\vec r_2)]\right\}.
\label{sigma3}
\end{equation}
The quark-photon distribution function in (\ref{DM}) reads,
\begin{equation}
\Psi_{\gamma q}(\alpha,\vec r)=
\frac{\sqrt{\alpha_{em}}}{2\pi}\,
\vec{e^*}
\chi_f 
\left\{ im_q\alpha^2\left[\vec n\times\vec\sigma\right]+
\alpha\left[\vec\sigma\times\vec\nabla\right]-i(2-\alpha)\vec\nabla
\right\}
\chi_i
K_0(\alpha m_q r),
\label{psi1}
\end{equation}
where $\chi_{i,f}$ are the quark spinors.

The $\bar qq$ dipole-nucleon cross section $\sigma_{\bar qq}(r)$ in (\ref{sigma3}) at a soft scale has been parametrized in a saturation form and  fitted to DIS and photoproduction data from NMC and HERA. The details can be found in \cite{kst2,applied}.

The quark distribution function in the proton at a hard scale is related to the structure function $F_2(x,Q^2)$, which is well known from DIS data, but only at a hard scale. At a soft scale one should rely on models, and we employ the well developed quark-gluon string model (QGSM) \cite{qgsm1}, or a similar dual parton model \cite{Capella}. Both models assume Regge behavior at the end-points $x\to 1$ or , $x\to 0$ of the quark distribution functions, and a simple, but ad hoc, interpolation at medium x. We skip the simple, but lengthy expressions. The details can be found e.g. in \cite{qgsm2}.

\begin{figure}[h]
    \centering
    \includegraphics[width=6.5cm]{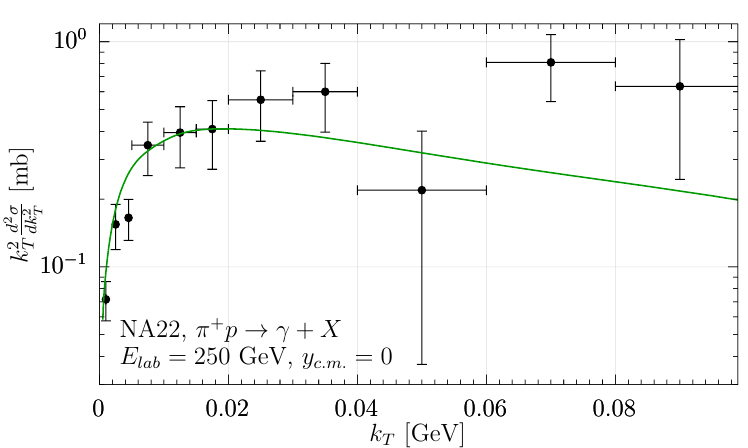}
    \hspace{1cm}
    \includegraphics[width=6.5cm]{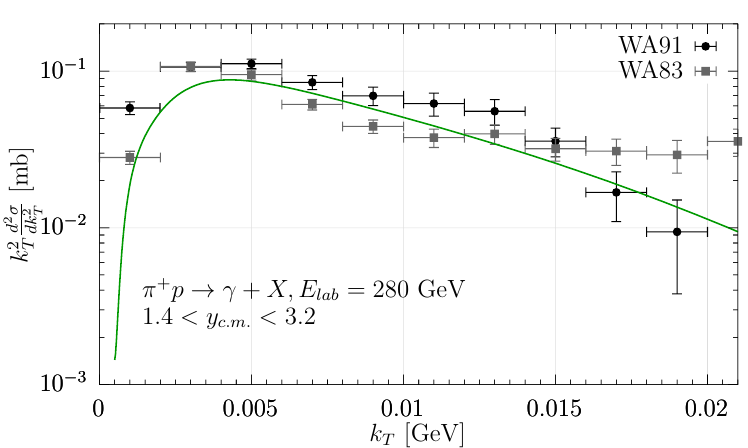}
    \caption{{\it Left:} Comparison with data of the NA22 experiment \cite{NA22} for $\pi^+p\to\gamma X$ at  $E_{lab}=250 GeV$. {\it Right:} Comparison with data of the WA91 \cite{WA91} and WA83 \cite{WA83} experiments for $\pi^+p\to\gamma X$ at  $E_{lab}=280 GeV$.}.
    \label{fig:fig1-2}
\end{figure}

\begin{wrapfigure}{l}{0.5\textwidth}
    \centering
    \includegraphics[width=6.5cm]{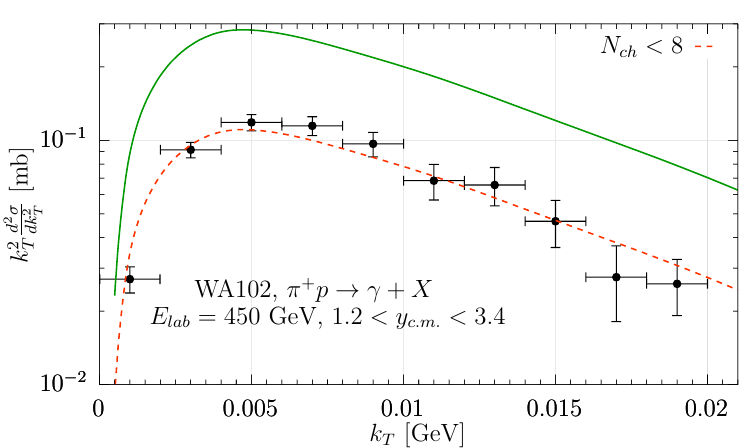}
    \caption{Comparison with data of the WA102 \cite{WA102} experiment for $\pi^+Be\to\gamma X$ at  $E_{lab}=450 GeV$.}
    \label{fig:fig3}
\end{wrapfigure}

At somewhat higher energy $E_{lab} = 450 GeV$ our calculations depicted by solid curve in Fig.~\ref{fig:fig3}, apparently overestimate the data of the WA102 experiment.
However, the experiment had specific cuts, namely, events with number of charge tracks $N_{ch} > 8$, were excluded. To calculate the multiplicity distribution we assume the Poisson distribution of the number of unitary cut Pomerons, and employed the result of QGSM. So we obtained a suppression factor $\delta=\sum\limits_{N_{ch}=0}^8/\sum\limits_{N_{ch}=0}^\infty=0.39$.

 Summarizing:

- Low’s paper considered only a large rapidity gap process of diffractive photon radiation $h\to h\gamma$, which has little to do with multi-particle production.

- Radiation of small-$k_T$ photons takes long time $\sim 1/k_T$ , so they cannot appear momentarily from the interaction blob, as is assumed in the BM.

- Due to electromagnetic neutrality of the Pomeron photon radiation at the mid-rapidities cancels. This interference effect is missed in BM.

 - The observed enhancement of low-$k_T$ photons in comparison with incorrect BM calculations, should not be treated as a puzzle.
 
- Calculations based on the color dipole description of photon radiation well agree with data.\\

\noindent
{\bf Acknowledgements:} 
We are thankful to Michal Krelina and Klaus Reygers
for numerous informative discussions.
 This work was supported in part by grants ANID - Chile FONDECYT 1231062 and 1230391,  by  ANID PIA/APOYO AFB220004, and by ANID - Millennium Science  Initiative Program 
ICN2019\_044.

\end{document}